\documentclass[10pt]{revtex4}
\usepackage{amssymb}
\usepackage{latexsym}
\usepackage{epsfig}
\newcommand{\nn}{\nonumber}

\newcommand{\ve}{\varepsilon}

\begin{document}                             

\title{Non-Linearly Interacting Ghost Dark Energy in Brans-Dicke Cosmology}

\author{E. Ebrahimi
\footnote{eebrahimi@mail.uk.ac.ir}$^{,a,b}$
,H.Golchin\footnote{h.golchin@.uk.ac.ir}$^{,a}$}
\address{$^a$ Department of Physics, Shahid Bahonar University, PO Box 76175, Kerman, Iran\\
$^b$Research Institute for Astronomy and Astrophysics of Maragha
(RIAAM), Maragha,
         Iran}

\begin{abstract}
In this paper we extend the form of interaction term into the
non-linear regime in the ghost dark energy model. A general form
of non-linear interaction term is presented and cosmic dynamic
equations are obtained. Next, the model is detailed for two
special choice of the non-linear interaction term. According to
this the universe transits at suitable time ($z\sim 0.8$) from
deceleration to acceleration phase which alleviate the coincidence
problem. Squared sound speed analysis revealed that for one class
of non-linear interaction term $v_s^2$ can gets positive. This
point is an impact of the non-linear interaction term and we never
find such behavior in non interacting and linearly interacting
ghost dark energy models. Also statefinder parameters are
introduced for this model and we found that for one class the
model meets the $\Lambda CDM$ while in the second choice although
the model approaches the $\Lambda CDM$ but never touch that.
\end{abstract}

 \maketitle
\section{Introduction}

QCD ghost dark energy (GDE) is one of models presented to solve
the unexpected acceleration problem of the
universe\cite{Rie,perl1,perl2,hanany,spergel,coll,teg,cole,springel}.
This model is based on the Veneziano ghost scalar field which was
presented to solve the so-called $U(1)$ problem \cite{veneziano}.
Taking a flat Minkowski background into account the ghost scalar
filed does not show any physical effect to vacuum energy of
spacetime. However, considering a curved  or dynamic spacetime
cancelation of the vacuum energy density will not be complete and
there remains a small contribution to the energy density. Authors
of \cite{urban,ohta} showed that in an expanding background this
contribution can be written as $\Lambda^3_ {QCD} H$, where H is
the Hubble parameter and $\Lambda^3_ {QCD}$ is QCD mass scale.
Taking $\Lambda_{\rm QCD}\sim 100 MeV$ and $H \sim 10^{-33}eV$
,$\Lambda^3_{\rm QCD}H\sim (3\times10^{-3}eV)$ which can alleviate
the fine tuning problem\cite{ohta}. Beside this nice resolution of
the fine tuning problem the GDE model is interesting because there
is no need to add a new degree of freedom to current state of the
physics literature. According to this form of energy density
,$\rho_{GDE}\sim H$, the ghost dark energy was proposed to solve
acceleration problem. Cai et.al considered the cosmological
consequences of this model of dark energy in \cite{CaiGhost}. They
constrained the parameters by observations and also discussed the
cosmic dynamics. Also authors considered $v_s^2$ analysis and
found signs of instability in the model. In \cite{ebrins}, the
authors considered $v_s^2$ analysis in details in presence and
absence of the non-gravitational interaction which the results
were in agreement with that of Cai et.al. Other features of the
GDE energy are investigated. Sheykhi et.al presented a tachyon
reconstruction of the model in \cite{shemov}. In this paper the
authors showed that the evolution of the ghost dark energy
dominated universe can be reconstructed by a tachyon scalar field.
The potential of the tachyon field are presented according to the
GDE model. Quintessence relation to the GDE is also considered in
\cite{sheykhi1}. A K-essence reconstruction of GDE is presented by
A. R. Fernandez \cite{fernandez}. The author discussed the kinetic
k-essence function F(X) in a flat background.

Interacting models of dark energy, introduced in the literature by
Wetterich in \cite{wett}. They found a solution to the so-called
"coincidence problem" which simply asks that why the DE component
becomes the dominant component at present time? The interaction
between DM and DE energy can be considered observationally and
theoretically. There are evidences which confirm a better
consistency between observations and interacting models of DE in
comparison with the non-interacting models. Examples are presented
in \cite{interact1,oli}. Also theoretically there is no reason
against interaction between dark sector components. If one(or
both)dark component does not interact with the rest of the
universe then it would be conserved. What symmetry in system
support such a conserved quantity? This question at present time
due to lack of a underlying theory for DM and DE does not receive
any answer. Then there is a lot of interest to models of DE which
leave a chance of interaction between dark sectors of the
universe. However, the form of interaction term is still a matter
of choice and generally can be written as $Q\sim\Gamma
f(\rho_m,\rho_D)$. In simplest level, $f$ can be taken as a linear
function of energy densities. The choice $f=\rho_m, \rho_D,
\rho_m+\rho_D$ is investigated widely in the literature
\cite{Ame,Zim,wang1}. Also authors tried to extend the form of
interaction term to non-linear regime. For example authors of,
\cite{jian}, showed that a product coupling, i.e., an interaction
term which is proportional to the product of $\rho_D$, $\rho_m$
and $\rho_D+\rho_m$ can be consistent with observations. Next in
\cite{arevalo}, the authors tried a general form of non-linear
interaction term and studied the cosmic dynamic of the universe in
presence of such interaction term. For a special case of the
non-linear interaction term they found analytic solutions which is
consistent with the supernova type Ia (SNIa) data from the Union2
set. Another example can be seen in \cite{BD3} which the authors
study DE models in presence of a non-linear interaction term from
a dynamical system point of view. Our aim in this paper is to
examine such non linear interaction term in BD framework in a two
component universe filled with GDE and cold dark matter.

The Brans-Dicke theory of gravity was proposed in 1961 \cite{BD}.
This model modifies the Einstein's gravity in a way admitting the
so called "Mach's principle". To this end , a new scalar degree of
freedom is added to incoroporate the Mach's principle. The
Brans–Dicke theory has passed experimental tests in the solar
system \cite{bertotti}. One can find more features of BD cosmology
in\cite{BD1,BD2}.

We also find the statefinder parameters, in order to discriminate
the model from the other dark energy models. The statefinder pair
parameters $\{r, s\}$, introduced by Sahni et al
\cite{Sahni:2002fz, Alam:2003sc}, is the geometrical diagnostic
pair, which is constructed from the scale factor and its
derivatives up to third order. This is a geometrical pair since it
is constructed from the space-time metric directly. Hence,
statefinder is more universal parameter to study the dark energy
models than any other physical parameters. It is possible to
construct a two dimensional space spanned by $r$ and $s$ which
different trajectories in this plane corresponds to different dark
energy models. Statefinder analysis for loop quantum cosmology and
$f(T)$ gravity can be find respectively in \cite{SF1,SF2}. In
\cite{SF3}, the statefinder parameters are constrained by
observational data.

The rest of this paper is organized as follows. In the next
section, we present the GDE model in the flat BD theory with a
closed form of interaction ($Q$). The next section is devoted to
the non-linearly interacting GDE model. In section \ref{r-s}, we
discussed the statefinder analysis for non-linearly interacting
model. We summarize our results in section \ref{sum}.

\section{Ghost dark energy in BD theory with linear interaction}\label{flat}

The action of the Brans-Dicke gravity with one scalar field $\phi$
can be written as
\begin{equation}
I_{G}=-\frac{1}{16\pi}\int_{\mathcal{M}}
d^{n+1}x\sqrt{-g}\left(\phi {R}-\frac{\omega}{\phi}(\nabla\phi)^2
+{\cal L}(m)\right),\label{act1}
\end{equation}
where ${R}$ is the scalar curvature, $\cal L$ is the Lagrangian
density of the matter and $\omega$ is the BD constant. Governing
field equations can be obtained by varying the action with respect
to $g_{\mu\nu}$ and $\phi$. The results read
\begin{eqnarray}
&&\phi \,G_{\mu\nu}=-8\pi T_{\mu\nu}^{M} - \frac{\omega}{\phi}
\left(\phi_{,\mu}\phi_{,\nu}-\frac{1}{2}g_{\mu\nu}\phi_{,\lambda}\phi^{,\lambda}\right)
-\phi_{;\mu;\nu}+g_{\mu\nu}\Box\phi, \label{eq1}
\\\nonumber & &\Box\phi=\frac{8\pi}{(n-1)\omega+n}T_{\lambda}^{M
\,\lambda}. \label{bdfeq}
\end{eqnarray}

One important point about BD theory is its varying gravitational
constant $G$. In BD gravity the, $G^{-1}_{\mathrm{eff}}={2\pi
\phi^2}/{\omega}$,  where $G_{\mathrm{eff}}$ is the effective
gravitational constant as long as the scalar field $\phi$ changes
slowly.

Taking the  Friedmann-Robertson-Walker (FRW) metric and above
equations into account we obtain the BD cosmology field equations
as
\begin{eqnarray}
 &&\frac{3}{4\omega}\phi^2\left(H^2+\frac{k}{a^2}\right)-\frac{1}{2}\dot{\phi} ^2+\frac{3}{2\omega}H
 \dot{\phi}\phi=\rho_M+\rho_D,\label{FE1}\\
 &&\frac{-1}{4\omega}\phi^2\left(2\frac{{\ddot{a}}}{a}+H^2+\frac{k}{a^2}\right)-\frac{1}{\omega}H \dot{\phi}\phi -\frac{1}{2\omega}
 \ddot{\phi}\phi-\frac{1}{2}\left(1+\frac{1}{\omega}\right)\dot{\phi}^2=p_D,\label{FE2}\\
 &&\ddot{\phi}+3H
 \dot{\phi}-\frac{3}{2\omega}\left(\frac{{\ddot{a}}}{a}+H^2+\frac{k}{a^2}\right)\phi=0,
 \label{FE3}
\end{eqnarray}
where $a(t)$ is the scale factor, and $k$ is the curvature
parameter with $k = -1, 0, 1$ corresponding to open, flat, and
closed universes, respectively. Also $H=\dot{a}/a$, is the Hubble
parameter, $\rho_D$ and $p_D$ are, the energy density and pressure
of the dark energy respectively and $\rho_M$, is the dark matter
density while we take a pressureless dark matter ($p_M=0$).

The energy conservation equations for DE and pressureless DM in
presence of interaction read

\begin{eqnarray}
&&
\dot{\rho}_D+3H\rho_D(1+w_D)=-Q,\label{consq12}\\
&&\dot{\rho}_{M}+3H\rho_{M}=Q, \label{consm12}
\end{eqnarray}
where $w_D=\frac{p_D}{\rho_D}$ is the equation of state parameter
of dark energy and $Q$ denote the interaction between DE and DM
components. Summing above equations one obtains
\begin{eqnarray}
\dot{\rho}+3H(\rho+p)=0,\label{consqtot}
\end{eqnarray}
where $\rho=\rho_D+\rho_M$ and $p=p_D+p_M=p_D$. Above equations
shows that although the presence of interaction term allows
transition of energy between two components but the total energy
content is still conserved. It is worth to note that in this paper
we ignore the radiation contribution to the energy content due to
its smallness.

Here our aim is to consider an interacting version of GDE in a
flat background in the BD framework. The ghost dark energy density
is\cite{ohta}
\begin{equation}\label{GDE}
\rho_D=\Lambda_{\rm QCD}^3 H,
\end{equation}
where $\Lambda_{\rm QCD}$ is QCD mass scale. With $\Lambda_{\rm
QCD}\sim 100 MeV$ and $H \sim 10^{-33}eV$ , $\Lambda^3_{\rm QCD}H$
gives the right order of magnitude $\sim (3\times10^{-3}eV)$ for
the observed dark energy density \cite{ohta}.

Adding above relation to the BD cosmology
Eqs.(\ref{FE1}-\ref{FE3}),we still have a problem in analyzing the
system. The system of equations is not closed and we need to
introduce more relations. To this end, according to already
presented papers in the literature we consider a power law
relation between BD scalar field $\phi$ and scale factor as

\begin{equation}\label{phipl}
    \phi\propto a^{\varepsilon}.
\end{equation}
There exist strong observational constraint on $\varepsilon$. Wu
and Chen in \cite{wu} using WMAP and SDSS data evaluated the rate
of change in the gravitational constant $G$ and using this
parameter found that $\varepsilon<0.01$. Also in this paper they
presented a constraint on the BD constant $\omega$. They found
that $\omega<-120.0$ or $\omega>97.8$ \cite{wu}. A solar-system
probe in \cite{bertotti} reveals that $\omega>40000$ while a
cosmic scale survey shows that $\omega$ is smaller than 40000
\cite{acqui}. In another paper using local astronomical
observations put a  lower bound on $\omega$ \cite{Will}.

 Considering the ansatz (\ref{phipl}) one finds

\begin{equation}\label{phidot}
    \frac{\dot{\phi}}{\phi}=\varepsilon \frac{\dot{a}}{a}=\varepsilon H.
\end{equation}

Assuming above formula, the first Friedmann equation (\ref{FE1})
in a flat background becomes

\begin{eqnarray}
  (1-\frac{2\omega}{3}\varepsilon^2+2\varepsilon)&=&\frac{4\omega}{3\phi^2
  H^2}(\rho_D+\rho_M), \label{frid1}\\
  \Rightarrow \gamma&=&\Omega_D+\Omega_M.\label{frid1p}
\end{eqnarray}

In writing above equation we used the following definitions
\begin{eqnarray}
\Omega_M&=&\frac{\rho_M}{\rho_{\mathrm{cr}}}=\frac{4\omega\rho_M}{3\phi^2
H^2}, \label{Omegam} \\
\Omega_D&=&\frac{\rho_D}{\rho_{\mathrm{cr}}}=\frac{4\omega\rho_D}{3\phi^2
H^2}, \label{OmegaD} \\
\gamma&=&1-\frac{2\omega}{3}\varepsilon^2+2\varepsilon.
\label{gamadef}
\end{eqnarray}

Taking time derivative from (\ref{GDE}) we reach

\begin{equation}\label{eq1}
    \frac{\dot{\rho_D}}{\rho_D}=\frac{\dot{H}}{H}.
\end{equation}

 As we mentioned we would like to consider the GDE with non-linear
 interaction in a flat background ($k=\Omega_k=0$). Taking time
 derivative of (\ref{frid1}) and dividing both side by $H^2$ one
 finds a relation for $\frac{\dot H}{H}$. Also doing a little
 algebra with (\ref{consq12}), we get an equivalent for
 $\frac{\dot{\rho_D}}{\rho_D}$. Inserting these relations in
 (\ref{eq1}), one can find

\begin{equation}\label{wgen}
    w_D=\frac{\gamma}{2\gamma-\Omega_D}\left[\frac{2\varepsilon}{3}-1-\frac{2Q}{3H\rho_D}\right].
\end{equation}

Next, we turn to the deceleration parameter

\begin{eqnarray}\label{q}
q=-\frac{\ddot{a}}{aH^2}=-1-\frac{\dot{H}}{H^2} =
\varepsilon+\frac{1}{2}+\frac{3}{2}\frac{\Omega_D w_D}{\gamma}.
\end{eqnarray}
Replacing relation (\ref{wgen}) in above equation we get

\begin{eqnarray}\label{qgen}
q=\varepsilon+\frac{1}{2}+\frac{3\Omega_D}{2(2\gamma-\Omega_D)}
\left[\frac{2\varepsilon}{3}-1-\frac{2Q}{3H\rho_D}\right].
\end{eqnarray}
To give a complete set of equations of cosmic dynamics we should
also bring dark energy density parameter evolution equation. To
this end one can take a time derivative of (\ref{OmegaD}) and
after a little algebra reach

\begin{equation}\label{OmegaDev}
\frac{d\Omega_D}{d\ln a}
=\frac{3\Omega_D}{2}\left[1-\frac{\Omega_D}{2\gamma-\Omega_D}(1+\frac{2Q}{3H\rho_D})+\frac{2\varepsilon}{3}
\frac{2(\Omega_D-\gamma)}{2\gamma-\Omega_D}\right],
\end{equation}

In order to find GR limit of above equations we should set
$\varepsilon=0$ ($\omega\rightarrow\infty$) and $\gamma=1$. Taking
GR limit of $w_D, q$, evolution equation (\ref{OmegaDev}) and
setting $Q=3b^2H(\rho_D+\rho_m)$ we get

\begin{equation}\label{wgenli}
    w_D=\frac{-1}{2-\Omega_D}\left[1+\frac{2b^2}{\Omega_D}\right],
\end{equation}
\begin{eqnarray}\label{qgenli}
q=\frac{1}{2}-\frac{3\Omega_D}{2(2-\Omega_D)}
\left[1+\frac{2b^2}{\Omega_D}\right],
\end{eqnarray}

\begin{equation}\label{OmegaDevli}
\frac{d\Omega_D}{d\ln a}
=\frac{3\Omega_D}{2}\left[1-\frac{\Omega_D}{2-\Omega_D}(1+\frac{2b^2}{\Omega_D})\right],
\end{equation}

which exactly correspond to those obtained in \cite{sheykhi4}.
Having the relations (\ref{wgen}), (\ref{qgen}) and
(\ref{OmegaDev}) at hand we are ready insert any form of
interactions and find the subsequent impacts on cosmic dynamics.

\section{Interacting Ghost dark energy with generalized interaction term in flat BD
theory}\label{flat} Nature and origin of the dark energy and dark
matter is still unknown. Thus there are a lot of features which
usually are consequences of choices. One important example of this
area is the interaction problem between dark components. When we
model the recent acceleration of the universe through DE, it seems
that in some features we get better agreement with observations if
we consider an interaction between dark components. An example is
evidences from galaxy cluster Abell A586  which show better
agreement withe theoretical approach when there exist interaction
between DE and DM \cite{interact1}. From theoretical view there is
not any reason against interaction DM and DE. Thus many authors
consider interacting models of DE. In one branch people start the
the model with introduction of Lagrangians which leads to
interacting models. Instances can be find in \cite{samisuji}. So
there is enough motivations observationally and theoretically to
consider interacting models of DE($Q\neq 0$). Since there is not
any underlying theory to inform us about the form of interaction
term, then people usually consider this issue phenomenologically.
One more helping point come from dimensional analysis which says
that the interaction term should have a same dimension as energy
density dimension. In simplest level people consider a linear
relation proportional to $\rho_D$, $\rho_M$ or sum of them.
However there still are many choices to be examined in this
regard. There are examples of interacting models which the
interaction term contains products of different energy components
\cite{jian}. As we mentioned in the introduction our aim in this
note is to extend such ideas to a general case which includes more
variety of these models. The form of interaction read
\begin{equation}\label{formnli}
   Q=3Hb^2 \rho_D^{\alpha} \rho_m^{\beta}\rho^{\xi},
\end{equation}
where $\rho=\rho_m+\rho_D$ and $b^2$ is coupling constant factor.
The powers $\alpha$, $\beta$ and $\xi$ specify the form of the
interaction. Dimensional analysis constraint above equation
through
\begin{equation}\label{pqrcons}
    \alpha+\beta+\xi=1.
\end{equation}

This form of interaction  with above constraint includes a variety
choices in the literature. For example , for ($\alpha$; $\beta$;
$\xi$) = (1; 0; 0) one find $Q \propto \rho_D$ and for ($\alpha$;
$\beta$; $\xi$)= (0; 1;0) the case $Q\propto \rho_m$ is retrieved.
This form of interaction term leave a chance of analytic solutions
in some cases\cite{arevalo,oliverose}.

Taking the relation (\ref{pqrcons}), it can be seen that

\begin{equation}\label{formnli2}
    Q=3Hb^2 \rho_D^{\alpha} \rho_m^{\beta}\rho^{-\alpha-\beta+1}.
\end{equation}
To use above interaction term, we will rewrite it as
\begin{equation}\label{qomeganl}
\frac{2Q}{3H\rho_D}=2b^2
\gamma^{-\alpha-\beta+1}\Omega_D^{\alpha-1}\Omega_m^{\beta},
\end{equation}
where in last step we used Eqs.(\ref{frid1p}). In order to obtain
the cosmic dynamic equations with this interaction form we insert
above relation in equations (\ref{wgen}), (\ref{qgen}) and
(\ref{OmegaDev}). The result read

\begin{equation}\label{wgennl}
    w_D=\frac{\gamma}{2\gamma-\Omega_D}\left[\frac{2\varepsilon}{3}-1-2b^2
\gamma^{-\alpha-\beta+1}\Omega_D^{\alpha-1}\Omega_m^{\beta}\right],
\end{equation}
\begin{eqnarray}\label{qgennl}
q=\varepsilon+\frac{1}{2}+\frac{3\Omega_D}{2(2\gamma-\Omega_D)}
\left[\frac{2\varepsilon}{3}-1-2b^2
\gamma^{-\alpha-\beta+1}\Omega_D^{\alpha-1}\Omega_m^{\beta}\right],
\end{eqnarray}

\begin{equation}\label{OmegaDevnl}
\frac{d\Omega_D}{d\ln a}
=\frac{3\Omega_D}{2}\left[1-\frac{\Omega_D}{2\gamma-\Omega_D}(1+2b^2
\gamma^{-\alpha-\beta+1}\Omega_D^{\alpha-1}\Omega_m^{\beta})+\frac{2\varepsilon}{3}
\frac{2(\Omega_D-\gamma)}{2\gamma-\Omega_D}\right].
\end{equation}

It is worth to mention that for choice $\alpha=\beta=0$, all above
equations convert to the linear interaction relation which was
presented at the end of previous section. Above equations reveals
the impact of the non-linear interaction term on the cosmic
dynamics. However, here we also consider the issue of stability
against small perturbations in the background using the squared
sound speed ($v_s^2$) analysis. In present time the universe is in
a DE dominated stable phase then any sign of
instability,$(v_s^2<0)$, can challenge the model. To this end we
use
\begin{equation}\label{v2def2}
    v_s^2=\frac{dP}{d\rho}=\frac{\dot{P}}{\dot{\rho}}=\frac{\rho}{\dot{\rho}}\dot{w}+w,
\end{equation}
where $w$ is the effective EoS parameter. After a little algebra
one finds

\begin{eqnarray}
v_s^2&=&\frac{-4\Omega_D}{\big(( 2\gamma-\Omega_D )
^{2}(\gamma-\Omega_D)(-3{\gamma}^{-\beta-\alpha+1}{\Omega_D}^{\alpha-1}
( \gamma-\Omega_D) ^{\beta}\Omega_D
{b}^{2}+\Omega_D\,\epsilon+3\gamma-3\Omega_D ) \big)}\big[
\nn\\&\big[& \frac{3}{2}\Omega_D^{2\alpha-1}
\big((\frac{\beta}{2}+\frac{\alpha}{2}-\frac{3}{2})
{\Omega_D}^{2}-\gamma ( \beta+\frac{3\alpha}{2}-\frac{7}{2})
\Omega_D+\gamma^2 ( \alpha-2 ) \big)( ( \gamma-\Omega_D
)^\beta)^{2}{b}^{4} (\gamma^{-\beta-\alpha+1}
 )^{2}\big]+\nn\\&\big[& (\gamma-\Omega_D) \Omega_D^{\alpha-1}( \gamma-\Omega_D) ^{\beta}
 \big(((\frac{\beta}{2}+\frac{\alpha}{2}-\frac{3}{2})\epsilon-\frac{3\beta}{4}
-\frac{3\alpha}{4}+3 ) {\Omega_D}^{2}-\gamma (  (
\beta+\frac{3\alpha}{2}-2 )
\epsilon-\frac{3\beta}{2}-\frac{9\alpha}{4}+6 )
\Omega_D\nn\\&+&{\gamma}^{2} ( \epsilon \alpha-\frac{3\alpha}{2}+3
) \big) {b}^{2}{\gamma}^{-\beta-\alpha+1}\big]-\frac{1}{3}(
\gamma-\Omega_D
 )  (  ( -\frac{\epsilon}{2}+\frac{3}{2} ) {\Omega_D}^{2}-3\gamma
\Omega_D+{\gamma}^{2} ( \epsilon+\frac{3}{2} )  ) (
\epsilon-\frac{3}{2} )  \big].
\end{eqnarray}

In order to study the impact of the non-linear interaction, we
discus two non-linear choice of the model.

\subsection{$Q=3Hb^2\frac{\rho_D^2}{\rho}(\alpha=2,\beta=0)$}

\begin{figure}\epsfysize=7cm
{ \epsfbox{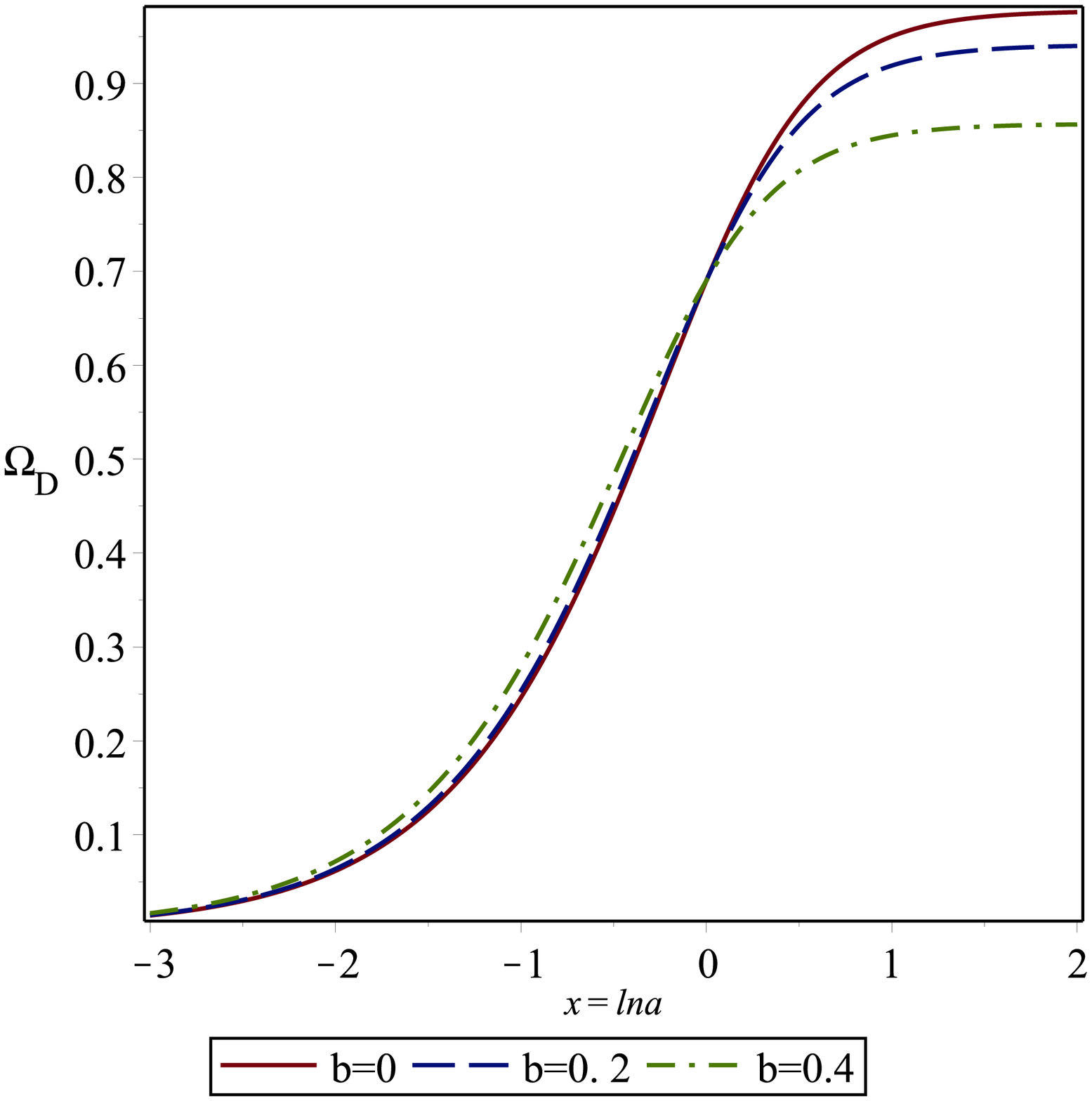}}\epsfysize=7cm {
\epsfbox{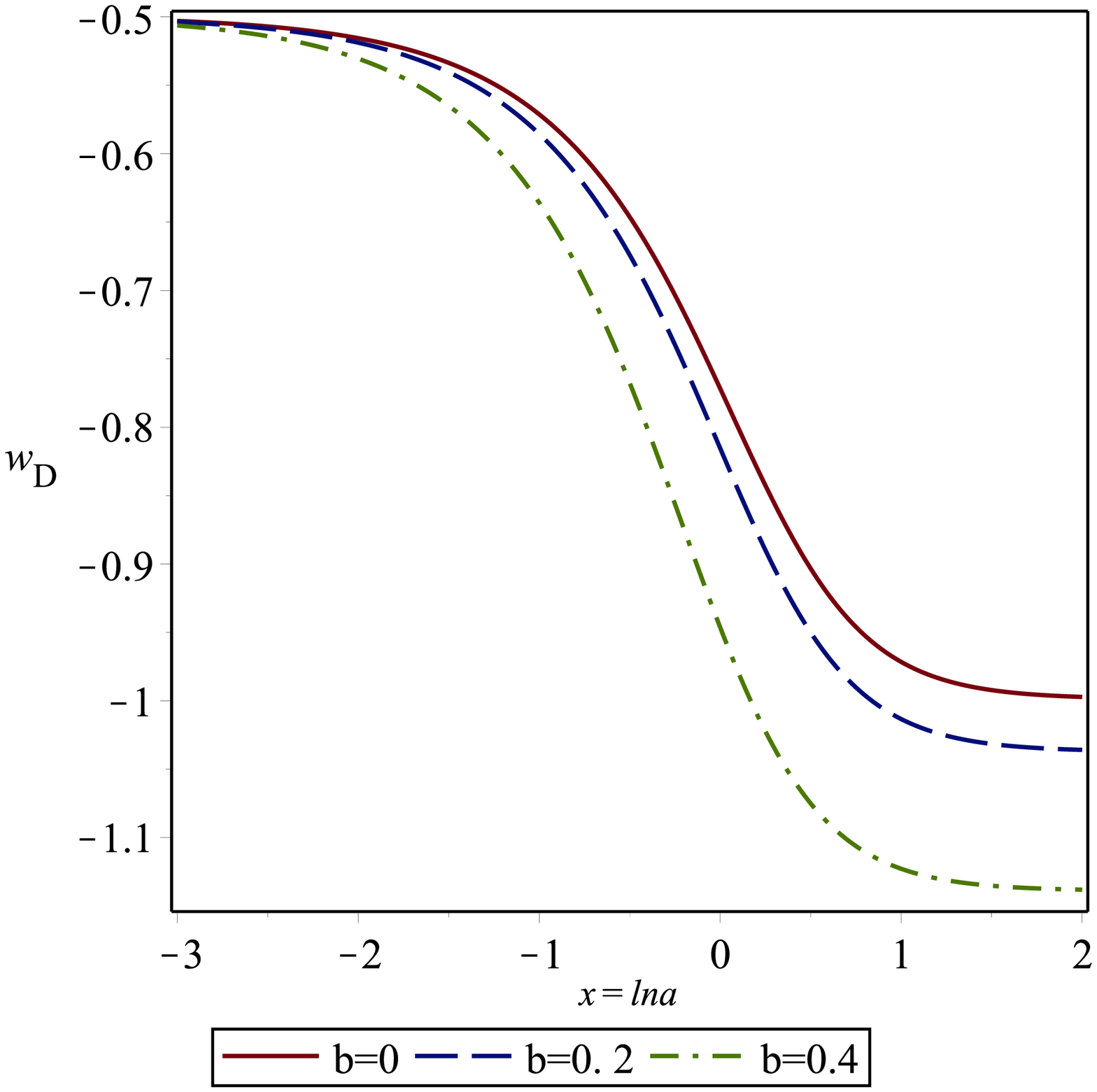}}\caption{In the left and right parts
$\Omega_D$ and $w_D$ are plotted against $x$ for coupling form
$Q=3Hb^2\frac{\rho_D^2}{\rho}$ respectively. In all figures
through the paper we set $a_0=1, \varepsilon=0.002$ and
$\omega=10000$} \label{i1}
\end{figure}

Plotting the density parameter equation versus the folding number
$x=\ln{a}$, it is obvious that $\Omega_D$ tends to constant value
rather than $1$. For larger $b$, more energy will be transferred
to the matter component and final value of $\Omega_D$ will be
smaller. However, for the DE component to catch the present value
$\Omega_{D0}=0.69$, there will be an upper limit on the coupling
constant $b$. One can see from Fig.(\ref{i1}), that the equation
of state parameter crosses the phantom barrier. It is also obvious
for this figure that for larger value of interaction parameter $b$
the EoS parameter takes more negative values and approaches a
limiting value at late time.

\begin{figure}\epsfysize=7cm
{ \epsfbox{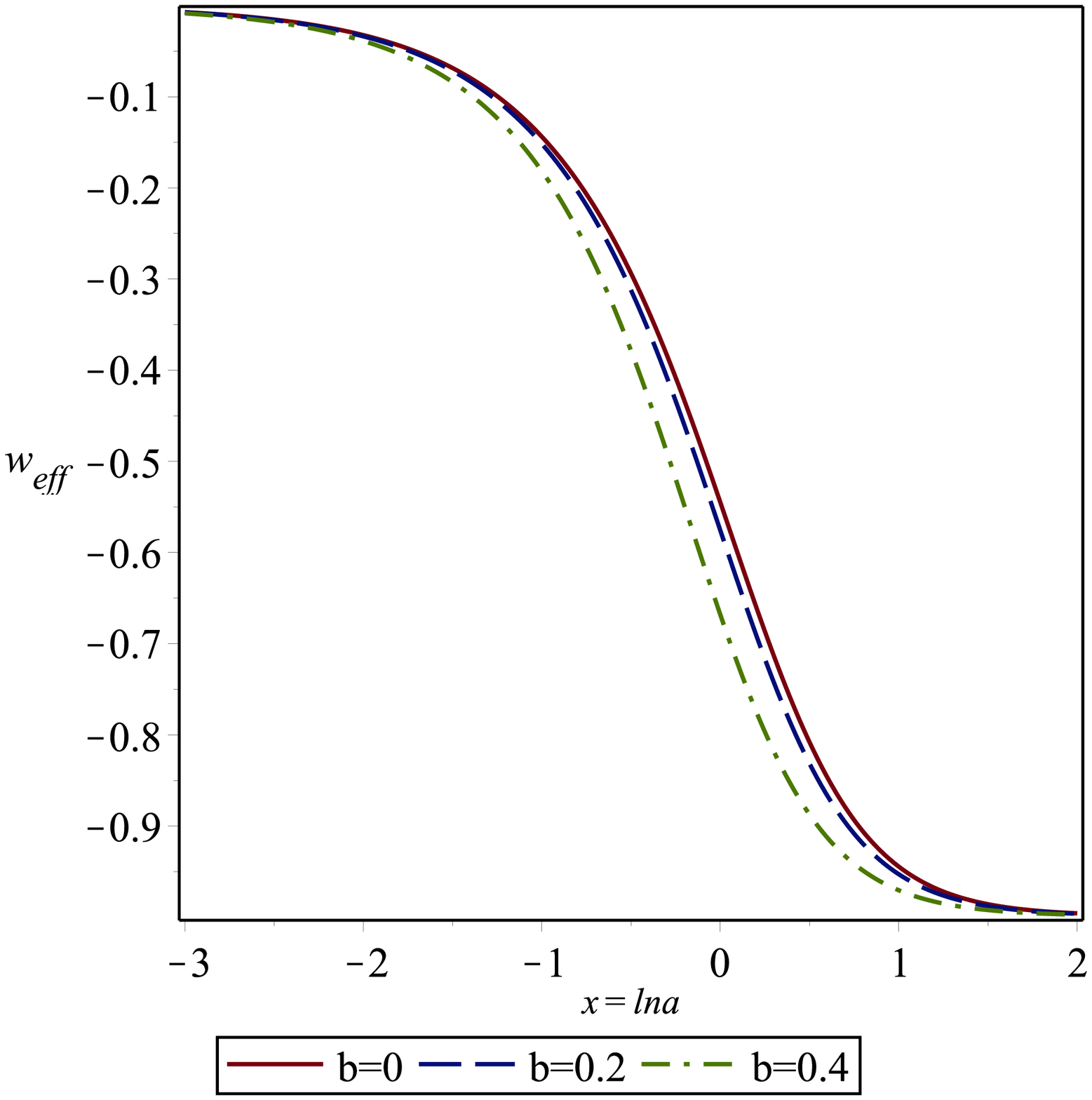}}\epsfysize=7cm {
\epsfbox{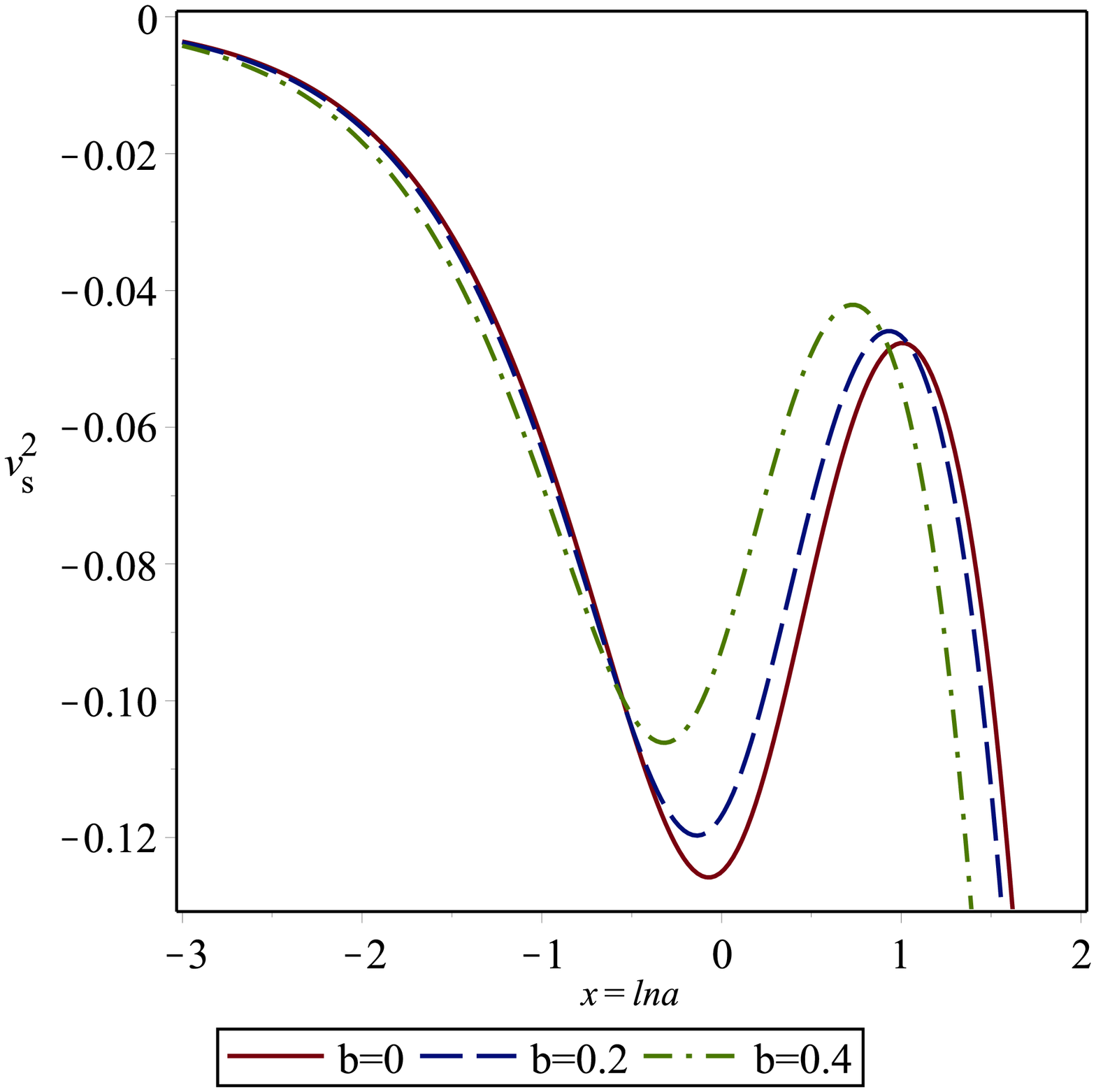}}\caption{$w_{eff}$ and $v_s^2$ for case
$(\alpha=2,\beta=0)$} \label{i2}
\end{figure}

Total fluid equation of state parameter is depicted in left part
of Fig.(\ref{i2}). This figure indicates that the total EoS
parameter approaches to $-1$ which reminds a big rip singularity
for destination of the universe. Also it can be seen from this
figure that in this model the universe enter the acceleration
phase earlier for larger $b$.

Finally, the squared sound speed is studied to check for
instability against small perturbations. $v_s^2$ is plotted
against $x$ in right part of figure (\ref{i2}). This figure
indicates that for all values of $b$, $v_s^2$ remains negative
which shows a sign of instability. So this model suffers the
stability problem although in other features the model seems to be
consistent with observations.

\subsection{$Q=3Hb^2\frac{\rho_D^3}{\rho^2}(\alpha=3,\beta=0)$}

\begin{figure}\epsfysize=7cm
{ \epsfbox{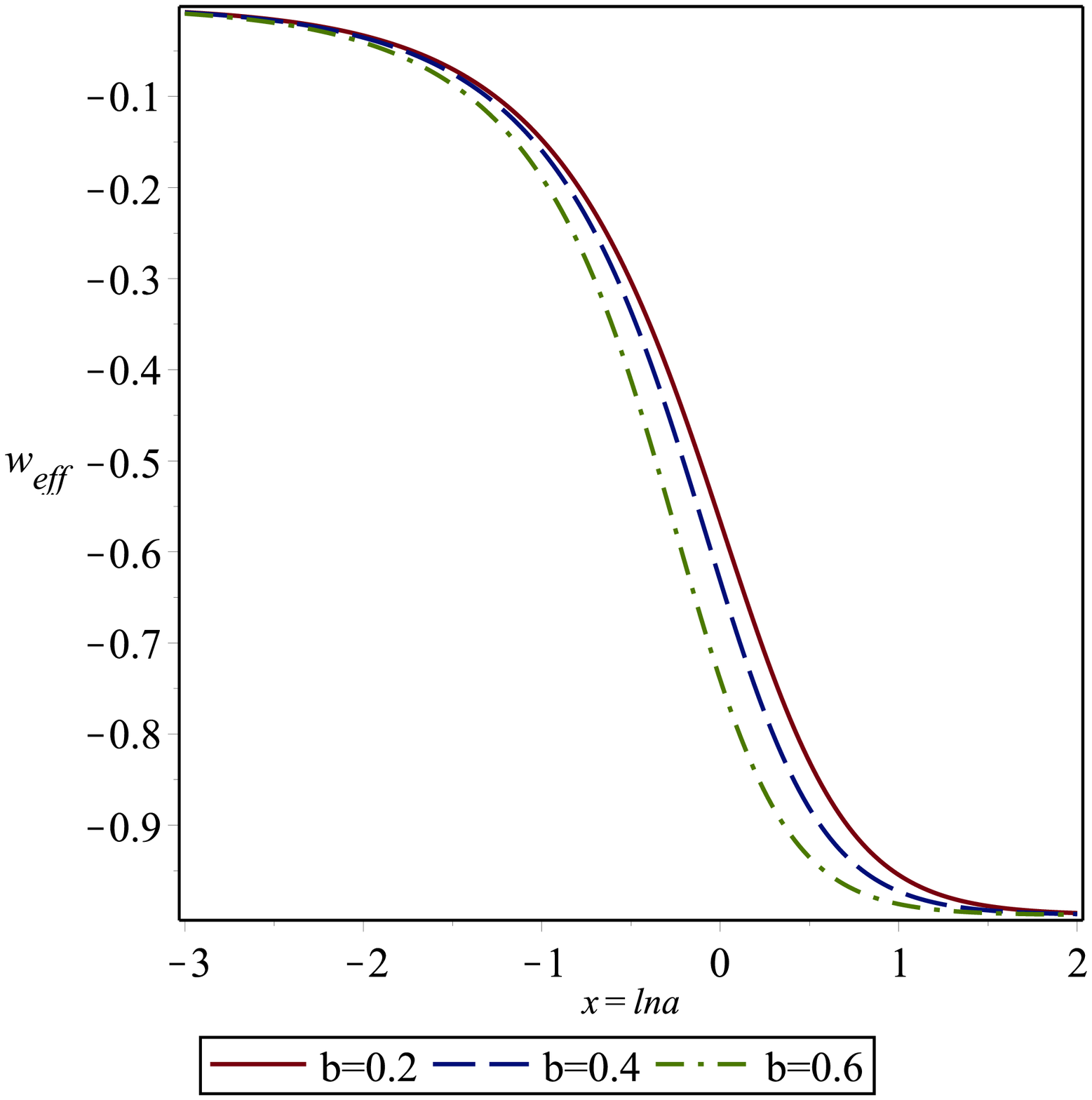}}\epsfysize=7cm {
\epsfbox{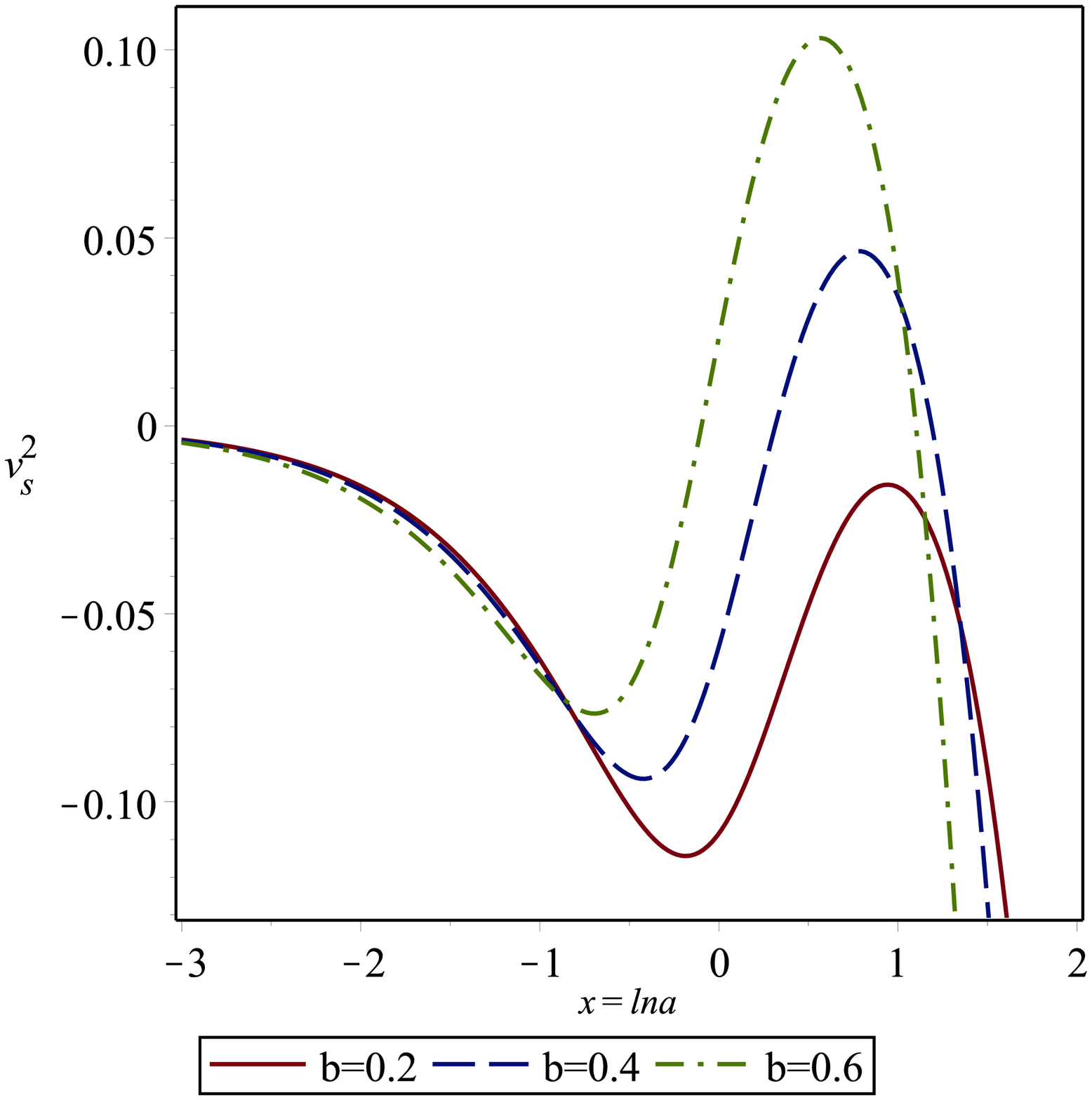}}\caption{$w_{eff}$ and $v_s^2$ for case
$(\alpha=3,\beta=0)$} \label{i3}
\end{figure}

As mentioned in the previous case all features of the model can be
consistent with observations but the squared sound speed analysis.
So we will try another choice of (\ref{formnli}) in this
subsection. Also in this case the EoS parameter will cross the
phantom line and approaches a constant value. Although the late
limit of $w_D$ in this case is smaller than that of the previous
case. In left part of Fig.(\ref{i3}), one can easily find that for
larger values of the coupling parameter $b$ the universe transits
to acceleration phase earlier. Also this figure indicates the
possibility of a big rip as fate of the universe because the final
limit of $w_{eff}$ is $-1$. Observationally, our universe seems to
transits from deceleration to acceleration at the redshift value
around $z>0.6$. It can be shown that for $b=0.5$ the universe
transition from matter to DE dominated phase will happen at
$x=-0.6(z\sim 0.8)$ which this point alleviate the coincidence
problem. As we presented in the previous interaction form, one
interesting point is the squared sound speed analysis which can
reveals signs of instability in the current dark energy dominated
universe. The previous case of interaction term was not able to
provide a positive $v_s^2$ region around present time. Since, In
present time the universe has stable, DE dominated phase of
expansion then we need to find models which there is no signs of
instability around present time. In the present form of
interaction term there exists periods of time which $v_s^2$ gets
positive values. The right part of Fig.(\ref{i3}), shows evolution
of the $v_s^2$ against $x=\ln{a}$. It seems from this figure that
for $b\geq0.5$ there is a period of time which $v_s^2$ achieves
positive values and after a finite era the $v_s^2$ will returns to
the negative domain. We also found that this point couldn't be
seen in linearly interacting or non-interacting models of GDE in
the BD framework and this is an important impact of the non-linear
interaction term.

\section{The statefinder diagnostic} \label{r-s}
To describe the evolution of the Universe, we use two cosmological
parameters H (the Hubble parameter) and q (the deceleration
parameter), However these two parameters can not differentiate
various dark energy models. In order to solve this problem, Sahni
et al \cite{Sahni:2002fz} have introduced a new geometrical
diagnostic pair parameter $\{r, s\}$, termed as statefinder. $r$,
$s$ are two dimensionless parameters which constructed from the
scale factor and its derivatives up to the third order as
\begin{equation} \label{rs}
r=\frac{\dddot a}{aH^3}\,,\qquad s=\frac{r-1}{3\big(q-1/2\big)}\,.
\end{equation}
$\{r, s\}$ are geometrical parameters since they constructed from
the cosmic scale factor alone, so the statefinder is more
universal than the physical variables which depend on the
properties of physical fields describing dark energy models. In a
flat $\Lambda$CDM model, the statefinder pair has fixed value
$\{r, s\} = \{1, 0\}$, also in the case of matter dominated
universe (SCDM) one finds $\{r, s\} = \{1, 1\}$\,.

Note that $r$ can be expressed in terms of the Hubble and the
deceleration parameters as \cite{Jawad:2014qoa} $r=\ddot
H/H^3-3q-2$. With the help of this equation and definition of the
$q$ in (\ref{q}), one can  rewrite $r$ as
\begin{equation} \label{rnew}
r=2q^2+q-\frac{\dot q}{H}\,.
\end{equation}
Noting (\ref{qgennl}) for the general form of interactions, one
finds that
\begin{equation} \label{qdot}
\dot q=\frac 32\,\frac{\dot \Omega_D (2\gamma-\Omega_D)+\dot
\Omega_D\,\Omega_D}{(2\gamma-\Omega_D)^2}\left(\frac{2\varepsilon}{3}-1-2b^2\gamma^{1-\alpha-\beta}\Omega_D^{\alpha-1}\Omega_m^{\beta}\right)-\frac{3\Omega_D}{2\gamma-\Omega_D}\left(b^2\gamma^{1-\alpha-\beta}(\alpha-1)\Omega_D^{\alpha-2}\dot
\Omega_D \Omega_m^{\beta}\right),
\end{equation}
where $\dot \Omega$ is given in (\ref{OmegaDevnl}). In the
following we discuss the statefinder for some non-linear
interactions between dark matter and the dark energy.

\vspace{3mm}
{\it case 1}): Linear interaction $Q=3Hb^2\rho$\,.\\
The linear interaction can be obtained by set $\alpha=3\,,
\beta=0$\, in (\ref{formnli}). In this case using
(\ref{qgennl})-(\ref{qdot}) one can find statefinder parameters as
\begin{eqnarray}
r&=&(2\gamma-\Omega_D)^{-3}\big[\big(16\ve^2+(24-48b^2)\ve+8+36b^4-36b^2\big)\gamma^3
-3\Omega_D\big(\frac 43\ve^2+(20-8b^2)\ve+7+9b^4-24b^2\big)\gamma^2\nn\\
&+&6\Omega_D^2\big(4+5\ve-6b^2+b^2\ve-\frac23\ve^2\big)\gamma-10\Omega_D^3\big],\\
s&=&\big[3(2\gamma-\Omega_D)^2(6b^2\gamma+3\Omega_D-4\ve\gamma)\big]^{-1}
\big[8(2\ve+3-3b^2)(-2\ve\!+3b^2)\gamma^3\!+6\Omega_D\big(\frac43\ve^2\!+(20-\!8b^2)\ve\!+\!3+9b^4\!-24b^2\big)
\gamma^2\nn\\
&-&12\Omega_D^2\big(b^2\ve-\frac23\ve^2-6b^2+3+5\ve\big)\gamma+18\Omega_D^3\big]\,.
\end{eqnarray}

\vspace{3mm}
{\it case 2}): $Q=3Hb^2\frac{\rho_D^3}{\rho^2}$\,.\\
In this case we should set $\alpha=3\,, \beta=0$\,, so  using
(\ref{qgennl})-(\ref{qdot}) one can find

\begin{eqnarray}
r&=&\big[\gamma(2\gamma-\Omega_D)\big]^{-3}\big[(16\ve^2+24\ve+8)\gamma^6-4\Omega_D
\big(\ve^2+15\ve+\frac{21}{4}\big)\gamma^5-4\Omega_D^2\big(\ve^2-\frac{15}{2}\ve-6\big)
\gamma^4\nn\\
&-&84\Omega_D^3\big(b^2\ve-\frac{3b^2}{14}+\frac{5}{42}\big)\gamma^3+78b^2
\Omega_D^4
\big(\ve-\frac{3}{26}\big)\gamma^2-12b^2\Omega_D^5\big(\ve+\frac34\big)\gamma-18b^4\Omega_D^6\big],\\
s&=&\big[3\gamma(2\gamma-\Omega_D)^2(6b^2\Omega_D^3+3\gamma^2\Omega_D-4\ve\gamma^3)\big]^{-1}
\big[(-32\ve^2-48\ve)\gamma^6+8\Omega_D\big(\ve^2\!+15\ve+\frac94\big)\gamma^5
+8\Omega_D^2\big(\ve^2\!-\frac{15}{2}\ve-\frac92\big)\gamma^4\nn\\
&+&168\Omega_D^3\big(b^2\ve-\frac{3b^2}{14}+\frac{3}{28}\big)\gamma^3-156b^2\Omega_D^4
\big(\ve-\frac{3}{26}\big)\gamma^2+24b^2\Omega_D^5\big(\ve+\frac34\big)\gamma+36b^4\Omega_D^6\big]\,.
\end{eqnarray}

\begin{figure}[tt]
\begin{picture}(0,0)(0,0)
\put(75,-10){\footnotesize Fig4a}
\put(225,-10){\footnotesize Fig4b} \put(395,-10){\footnotesize Fig4c} 
\end{picture}
\includegraphics[height=55mm,width=50mm,angle=0]{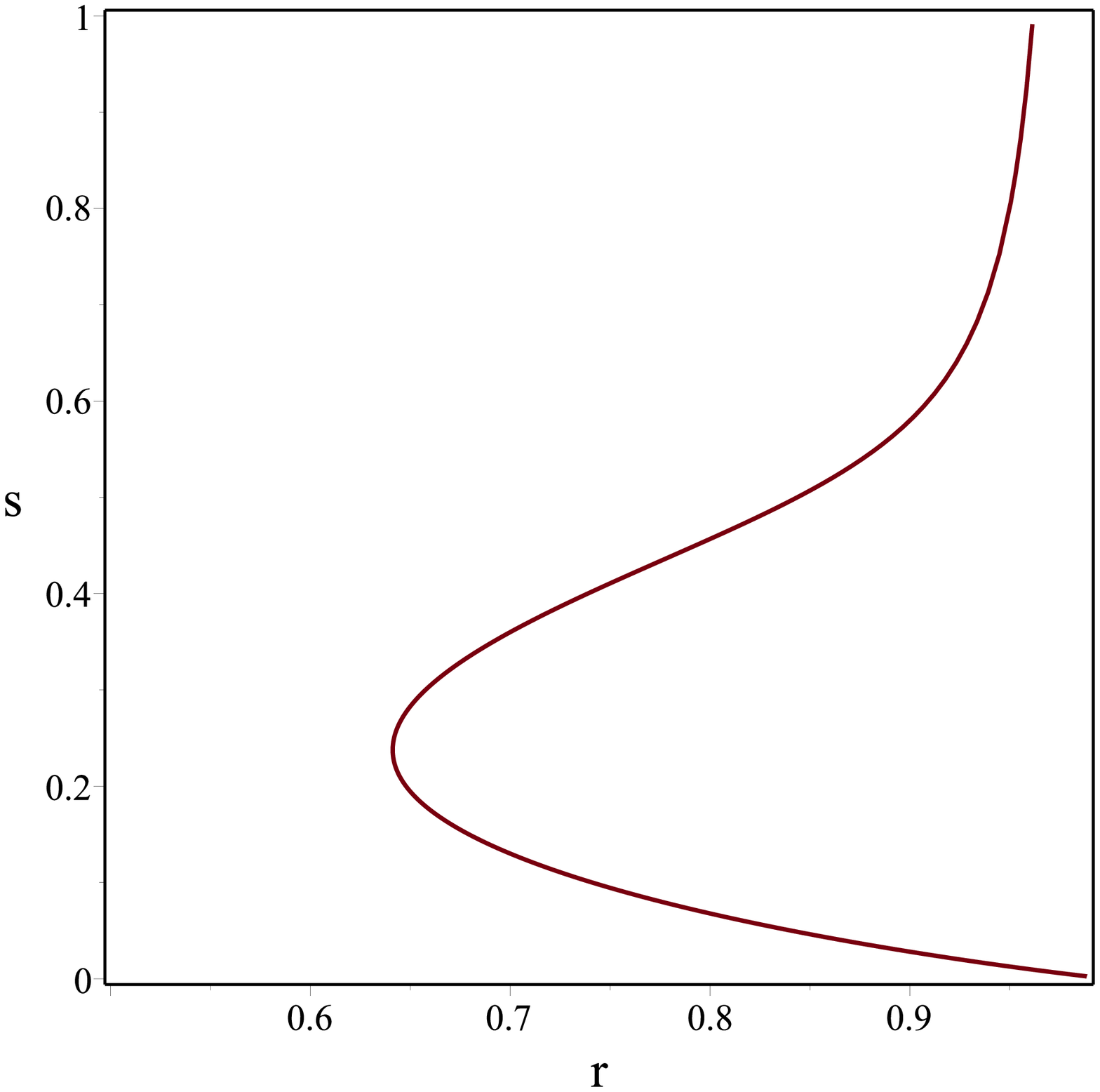} \,\,\,\,
\includegraphics[height=55mm,width=50mm,angle=0]{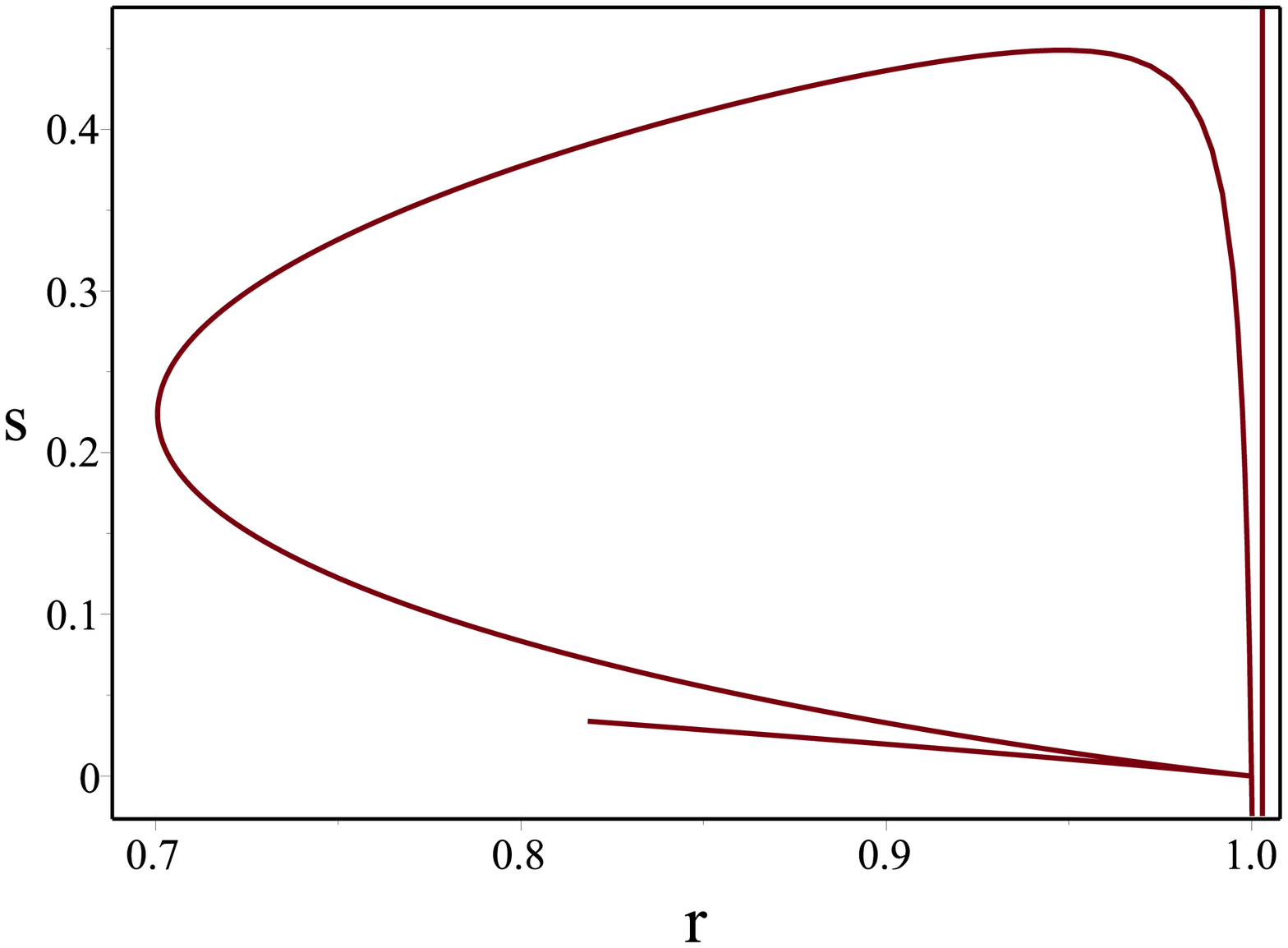} \quad \,\,
 \includegraphics[height=55mm,width=50mm,angle=0]{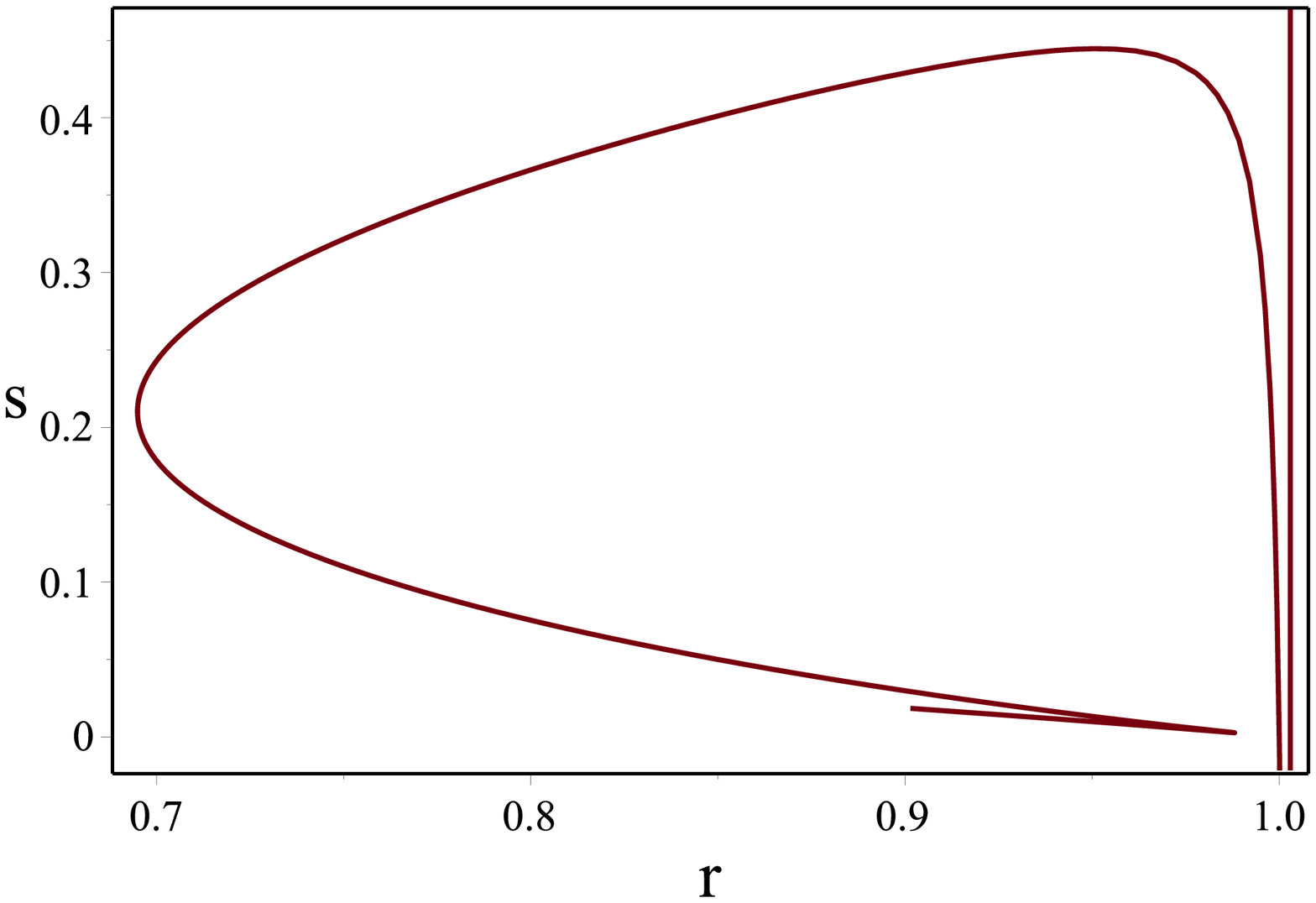}
\vspace{2mm} \caption{{\footnotesize The evolution of statefinder
parameter $r$ versus $s$ in the case of linear (Fig1a) and
non-linear interactions (Fig1b,c) where we set $\gamma=1$,
$\epsilon=.002$ and $b=0.1$\,. All curves lie in the region of the
quintessence models. In non-linear interaction of case 2 (Fig1b)
the trajectory cross the $\Lambda$CDM fixed point. In the case of
linear interaction and non-linear interaction of case 3 (Fig1c),
the trajectories tend but do not touch the $\Lambda$CDM fixed
point.}} \label{figrs}
\end{figure}

\vspace{3mm}
{\it case 3}): $Q=3Hb^2\frac{\rho_D^2}{\rho}$\,.\\
This non-linear interaction can be obtained by setting
$\alpha=2\,, \beta=0$\, in (\ref{formnli}). In this case repeating
(\ref{qgennl})-(\ref{qdot}) one can find statefinder parameters
(due to briefness, we do not demonstrate $r$, $s$ in this case). \\
In figure (\ref{figrs}) we have depicted the evolution of
$\{r,s\}$  for the above cases. This figure shows that statefinder
analysis can discriminate the models, where in all cases, curves
lie in the region $r<1, s>0$ which implies the quintessence
characteristic of the model \cite{Alam:2003sc}, as on expected. In
fig 4.b (non-linear interaction of case 2) the curve pass through
the $\Lambda$CDM fixed point $\{r=1,s=0\}$ but in fig 4.a and 4.c,
(linear interaction and non-linear interaction of case 3), the
trajectories tend to the to the $\Lambda$CDM fixed point but do
not touch it.

\section{Conclusion}\label{sum}
In different models of dark energy the the form of interaction
between dark matter and dark energy components has a linear
dependency to $\rho_m$ and $rho_D$. In the absence of an
underlying theory for DE and DM and also due to lack of
observational evidences in this field, one can try the possibility
of an interaction term with non-linear dependency to $\rho_m$ and
$\rho_m$. In this paper we considered the consequences of such
choices for interaction term in ghost dark energy model (GDE) in
the Brans-Dicke framework.

To this goal we considered a non-linear form of interaction as
$Q=3Hb^2 \rho_D^{\alpha} \rho_m^{\beta}\rho^{-\alpha-\beta+1}$ and
obtained the EoS parameter($w_D$), deceleration parameter ($q$)
and evolution equation of the density parameter for GDE in BD
framework in a flat background. These relations are presented for
a general choice of the interaction term (arbitrary $\alpha$ and
$\beta$). Next, we discussed two special choices of the
interaction term. The first choice is
$Q=3Hb^2\frac{\rho_D^2}{\rho}$. We obtained evolution of the
density parameter ($\Omega_D$) in this case versus $x=\ln a$.Using
this quantity we depicted the evolution of $w_D, w_{eff}$ and
$v_s^2$ versus $x$. The model sounds pretty well with what we
expect. For example shows a long deceleration phase which ends at
past and transits to a phase of acceleration which can solve the
coincidence problem. However this model suffers the stability
problem according to the squared sound speed ($v_s^2$) analysis.
In this model the squared sound speed is always negative and never
gets positive implying signs of instability against small
perturbations in the background. Due to this reason we make
another choice the interaction term to see if the ansatz of the
non-linear interaction term is capable to remove such problem.
Then we turn to the choice $Q=3Hb^2\frac{\rho_D^3}{\rho^2}$. In
this choice of the interaction term all good features of the model
are kept and the problem of negativity of the $v_s^2$ is removed.
For suitable choice of coupling parameter $b$ we obtained a
confined period of time which $v_s^2$ gets positive. However once
again $v_s^2$ will enter negative domain which this point is in
contrast with the same situation in the Einstein's gravity
\cite{ebnlgde}. We have to emphasis here that in a GDE model in BD
theory in non interacting case and also with linearly interacting
case we never find such period of time which $v_s^2$ gets positive
and this stage is an impact of the non-linear choice of the
interaction term.

The statefinder diagnostic is also presented in the next section.
In non-linear interaction of case
($Q=3Hb^2\frac{\rho_D^3}{\rho^2}$) the $\{r,s\}$ curve pass
through the $\Lambda$CDM fixed point $\{r=1,s=0\}$ but for case
$(Q=3Hb^2\frac{\rho_D^2}{\rho})$, the trajectory tend to the to
the $\Lambda$CDM fixed point but do not touch it.

It is worth to mention that although the latter choice seems more
consistent with what we expect a dependable model of DE but we
need to take closer look at different features discriminating the
model. Consistency with observational data and more subtle issues
for non-linearly interacting models of GDE is now under
investigation and will be addressed elsewhere.

\acknowledgments{This work has been supported financially by
Research Institute for Astronomy and Astrophysics of Maragha
(RIAAM) under project NO. 1/4165-54}

\end{document}